\documentclass[10pt,a4paper,twocolumn]{article}
\usepackage{titlesec}
\usepackage{color,soul}
\usepackage[utf8]{inputenc} 
\usepackage[affil-it]{authblk}
\usepackage{amsmath}
\usepackage{varioref}
\usepackage[colorinlistoftodos]{todonotes}

  \usepackage[pdftex]{hyperref} 
\let\OLDthebibliography\thebibliography
\renewcommand\thebibliography[1]{
  \OLDthebibliography{#1}
  \setlength{\parskip}{1pt}
  \setlength{\itemsep}{1pt plus 0.3ex}
}
\usepackage{siunitx}
\usepackage{float}
\usepackage{tikz}
\usepackage{xcolor}
\usepackage{comment}
\usepackage{braket}
\usepackage{nicefrac}
\usepackage[english]{babel} 
\usepackage[top=2.8cm,bottom=2.8cm,left=2.cm,right=2.cm]{geometry}  
\usepackage{amsmath,amsfonts,amssymb}  
\usepackage{caption}
\usepackage{graphicx}  
\graphicspath{{./Figures/}{./PSTricks/}}  
\usepackage{multirow}
\usepackage{appendix}
\usepackage{wrapfig}
\usepackage{setspace}
\usepackage{cite}
\usepackage{csquotes}
\usepackage{array,booktabs}

\usepackage{enumitem}
\usepackage{caption3} 
\DeclareCaptionOption{parskip}[]{}
\usepackage[font=small,labelfont=bf]{caption}
\usepackage[final]{pdfpages}
\usepackage[colorinlistoftodos]{todonotes}
\usepackage{tikz}
\usepackage{sidecap}
\sidecaptionvpos{figure}{t}
\usepackage{abstract}

\usepackage[export]{adjustbox}
\usepackage{dblfloatfix}
\begin{document}
%\maketitle
\title{\textbf{Field trial of a finite-key quantum key distribution system in the Florence metropolitan area}}
\author[1,*]{\small Davide Bacco}
\author[2,3]{\small Ilaria Vagniluca}
\author[1]{\small Beatrice Da Lio}
\author[3]{\small Nicola Biagi}
\author[4]{\small Adriano Della Frera}
\author[5]{\small Davide Calonico}
\author[3,6]{\small Costanza Toninelli}
\author[3,6]{\small Francesco S. Cataliotti}
\author[3,6]{\small Marco Bellini}
\author[1]{\small Leif K. Oxenløwe}
\author[3,6]{\small Alessandro Zavatta}

\affil[1]{\footnotesize CoE SPOC, DTU Fotonik, Department of Photonics Engineering, Technical University of Denmark, DK}
\affil[2]{\footnotesize Università degli Studi di Napoli "Federico II", Via Cinthia 21, 80126 Napoli, IT}
\affil[3]{\footnotesize CNR-INO, Istituto Nazionale di Ottica, Via Carrara 1, 50019 Sesto F.no, Firenze, IT}
\affil[4]{\footnotesize Micro Photon Devices S.r.l., Via Stradivari 4, 39100 Bolzano, IT}
\affil[5]{\footnotesize I.N.Ri.M. Istituto Nazionale di Ricerca Metrologica, Torino, IT}
\affil[6]{\footnotesize LENS and Università di Firenze, Via Carrara 1, 50019 Sesto F.no, Firenze, IT}
\affil[*]{dabac@fotonik.dtu.dk}

\date{} 
\pagestyle{plain}
\setcounter{page}{1}
\twocolumn[ 
\begin{@twocolumnfalse}
\maketitle
     \vspace{-0.8cm}
\begin{abstract}
\normalsize
\vspace*{-1.0em}
\noindent 
In-field demonstrations in real-world scenarios boost the development of a rising technology towards its integration in existing infrastructures. %Quantum key distribution (QKD) is an emerging quantum technology that still suffers from high costs and low performances, preventing it from a large-scale deployment in telecommunication networks.
Although quantum key distribution (QKD) devices are already adopted outside the laboratories, current field implementations still suffer from high costs and low performances, preventing this emerging technology from a large-scale deployment in telecommunication networks. Here we present a simple, practical and efficient QKD scheme with finite-key analysis, performed over a 21 dB-losses fiber link installed in the metropolitan area of Florence (Italy). Coexistence of quantum and weak classical communication is also demonstrated by transmitting an optical synchronization signal through the same fiber link.
\end{abstract}
  \end{@twocolumnfalse}
 ]
\section*{Introduction}
\vspace{-0.25cm}
In a society based on the continuous exchange of sensitive data and information, the importance of secure and trustful communications is essential. Quantum key distribution (QKD) allows to share data in an information-theoretical secure way, no longer based on computational assumptions but exploiting the basic principles of quantum mechanics~\cite{BB84,Scarani2009,Diamanti2016}. 
During the last 30 years, many QKD protocols have been developed and tested over optical fiber spools in laboratory demonstrations, achieving long transmission distances and key generation rates up to hundreds of Mbit/s~\cite{Boaron2018_421km,Hwang2009,Minder2019,Yin2016,Beatrice2018,Dynes2016}
However, this technology is still far from a large-scale deployment in existing fiber networks and telecom infrastructures, due to multiple factors: low secret-key rate, limited distance between users, lack of applications, high costs and high requirements in terms of low-noise fiber links.
In order to reveal practical controversies in real-world deployments, several QKD field trials have been implemented by exploiting installed fiber links on a metropolitan scale, with tens of kilometers of typical distance between nodes \cite{Qiu2014,Peev2014,Yuan2005,Shimizu2014,Tang2016,Bunandar2018,Collins2016,Zhang2017,Tanaka2008,Choi2014,Wonfor2017,Mao2018}. 
\begin{figure}[h!]
\centering
\includegraphics[width=0.47\textwidth]{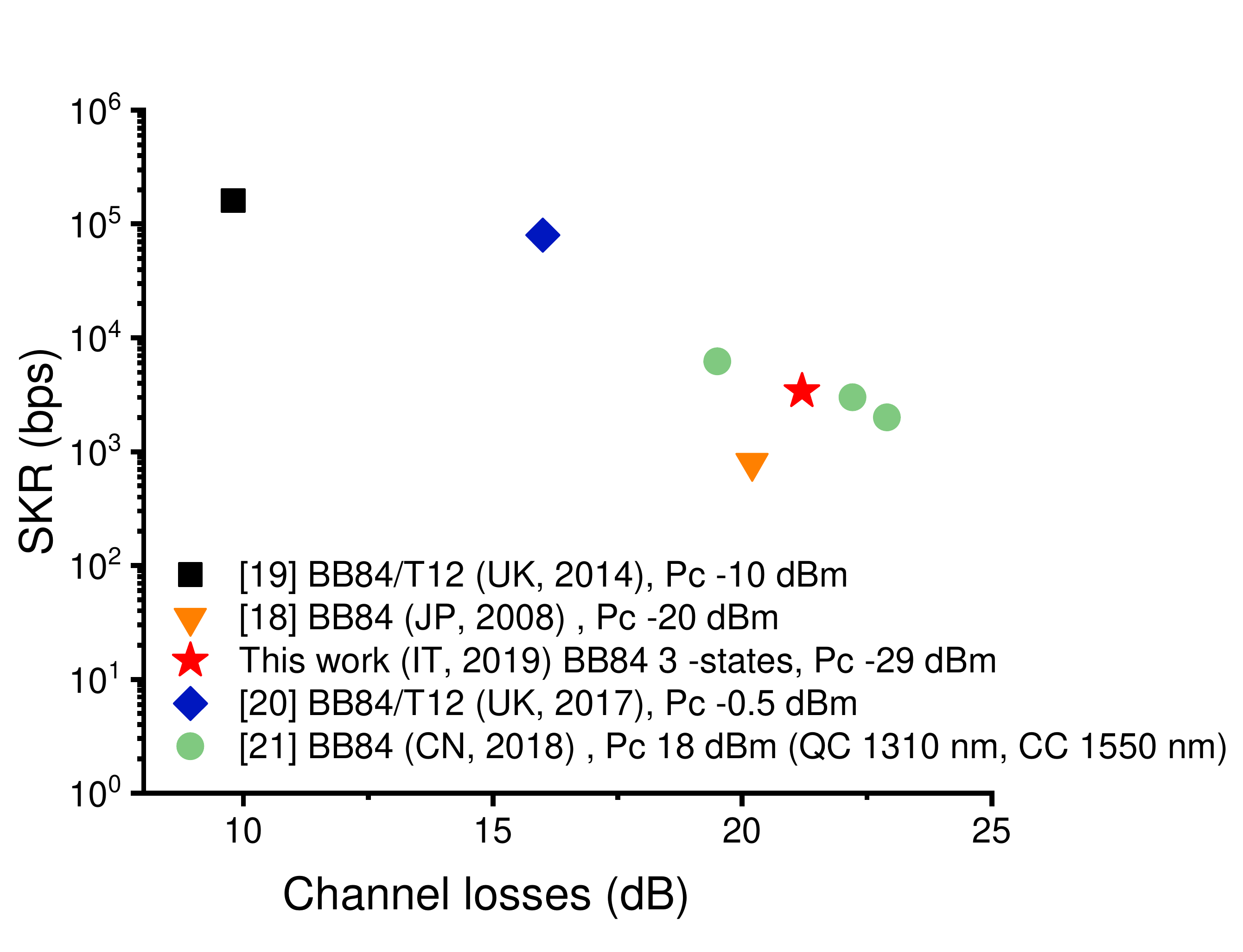}
\caption{{\bf State of the art of QKD field trials.} Recent QKD demonstrations with classical and quantum signal propagating through the same installed fiber. Total classical launch power is reported in dBm. Ref \cite{Tanaka2008} used SNSPDs detectors, all the other used InGaAs detectors.}
\label{fig:FieldTrialsSOTA}
\end{figure}
\begin{figure*}[h!]
\centering
\includegraphics[width=1\textwidth]{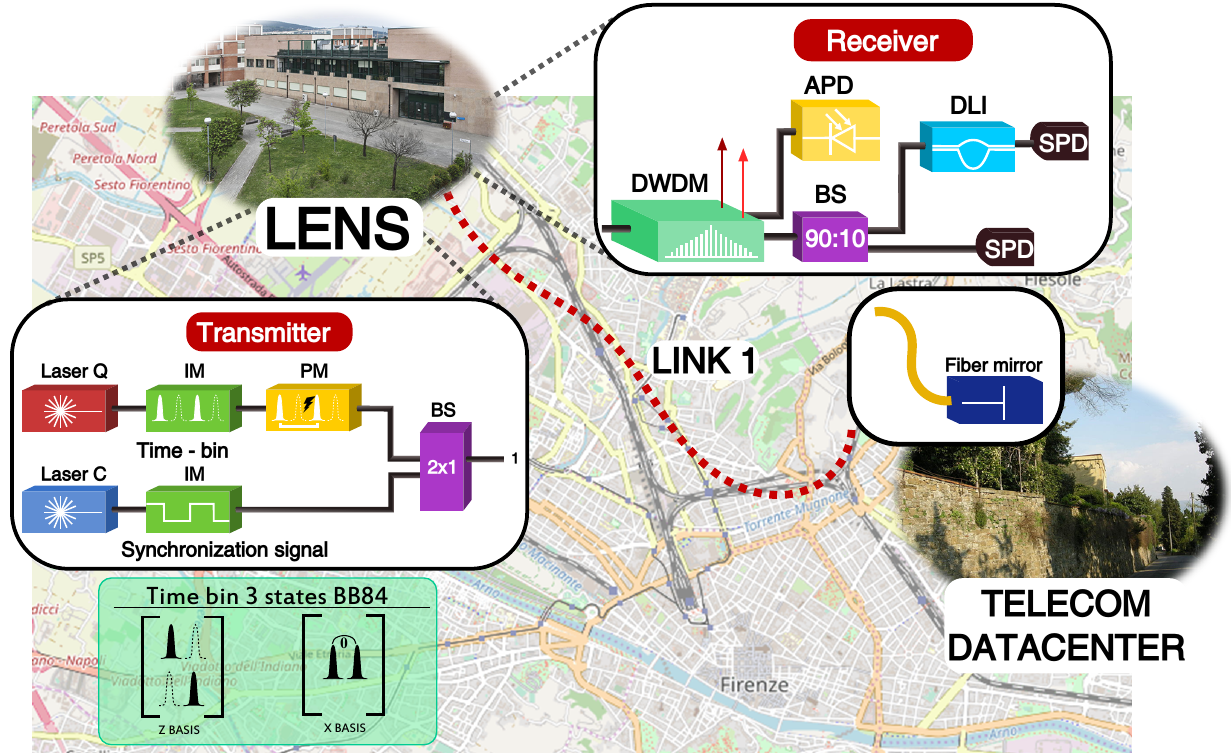}
\caption{{\bf Experimental setup} The transmission channel is a dark fiber link connecting LENS laboratory, where Alice and Bob are located, to a telecom datacenter, in a loop-back configuration. Laser Q: continuous wave laser at 1560.61 nm; Laser C: continuous wave laser at 1536.61 nm; IM: intensity modulator; PM: phase modulator; BS: beam splitter; DWDM: dense wavelength division multiplexing filter; DLI: delay line interferemoter; SPD: single photon detector. In the green square we reported the three quantum states prepared by Alice and propagated through the channel.}
\label{fig:setup}
\end{figure*}
Deployed commercial channels inherently suffer from higher losses and noise (due to splices, connections, bends and interfiber cross-talk), moreover their long-term stability is affected by changes in the environmental conditions and physical stress. Although many field trials were performed on a dark fiber (thus requiring a dedicated link for quantum key transmission), other experiments tested the coexistence between weak quantum signals and classical intense pulses propagating through the same installed fiber \cite{Tanaka2008,Choi2014,Wonfor2017,Mao2018}.
Classical light includes synchronization signals for QKD \cite{Tanaka2008} or high-speed data traffic \cite{Choi2014,Wonfor2017,Mao2018} and is mixed with quantum signals by means of dense wavelength division multiplexing (DWDM) schemes, often combined with polarization multiplexing \cite{Wonfor2017} and wide-range wavelength multiplexing (WDM) in C- and O-band \cite{Mao2018}. Recent demonstrations are presented in Figure \ref{fig:FieldTrialsSOTA}, where the average secret key rate achieved is reported with the corresponding transmission losses and total classical launch power.\\ 
%Recent field-demonstrations are presented in Figure \ref{fig:FieldTrialsSOTA}, where the average secret key rate achieved (SKR) is reported with the corresponding transmission losses. Many of these experiments were performed on a dark fiber, thus requiring a dedicated link for quantum key transmission only *cite*. Other field trials tested the coexistence between weak quantum signals and classical intense pulses propagating through the same fiber, by means of wavelength division multiplexing schemes in C-band. and O-band windows *cite*. Classical light transported through the same fiber include synchronization signals for QKD *cite* but also high-speed data traffic *cite*. The total launch power of classical signals, if present, is reported in Figure \ref{fig:FieldTrialsSOTA} for each field trial.\\
In this work we present a low-cost field demonstration of a complete QKD system, performed over an installed fiber link situated in Florence and exhibiting 21 dB transmission losses. A secret key rate of 3.4 kbit/s is achieved with a finite-key analysis, in the case of simultaneous transmission of synchronization signal (with -29 dBm launch power) and quantum signal through the same fiber at a different wavelength in the C-band. Time stability of the apparatus is demonstrated as well, by employing a servo-locked fiber-based interferometer for security checking.\\
The fiber link adopted for key exchange is a portion of a dark-fiber network connecting the entire Italian peninsula, from Turin National Institute of Metrological Research (INRIM) to Matera Space Center. This installed fiber about 1700 km in length, currently employed for time standard dissemination, constitutes the proper environment for a future setup of a large-scale quantum communication network, referred as the Italian Quantum Backbone \cite{INRIM2016}.

\section*{Experimental setup}
\begin{figure}[t]
      \includegraphics[width=0.5\textwidth]{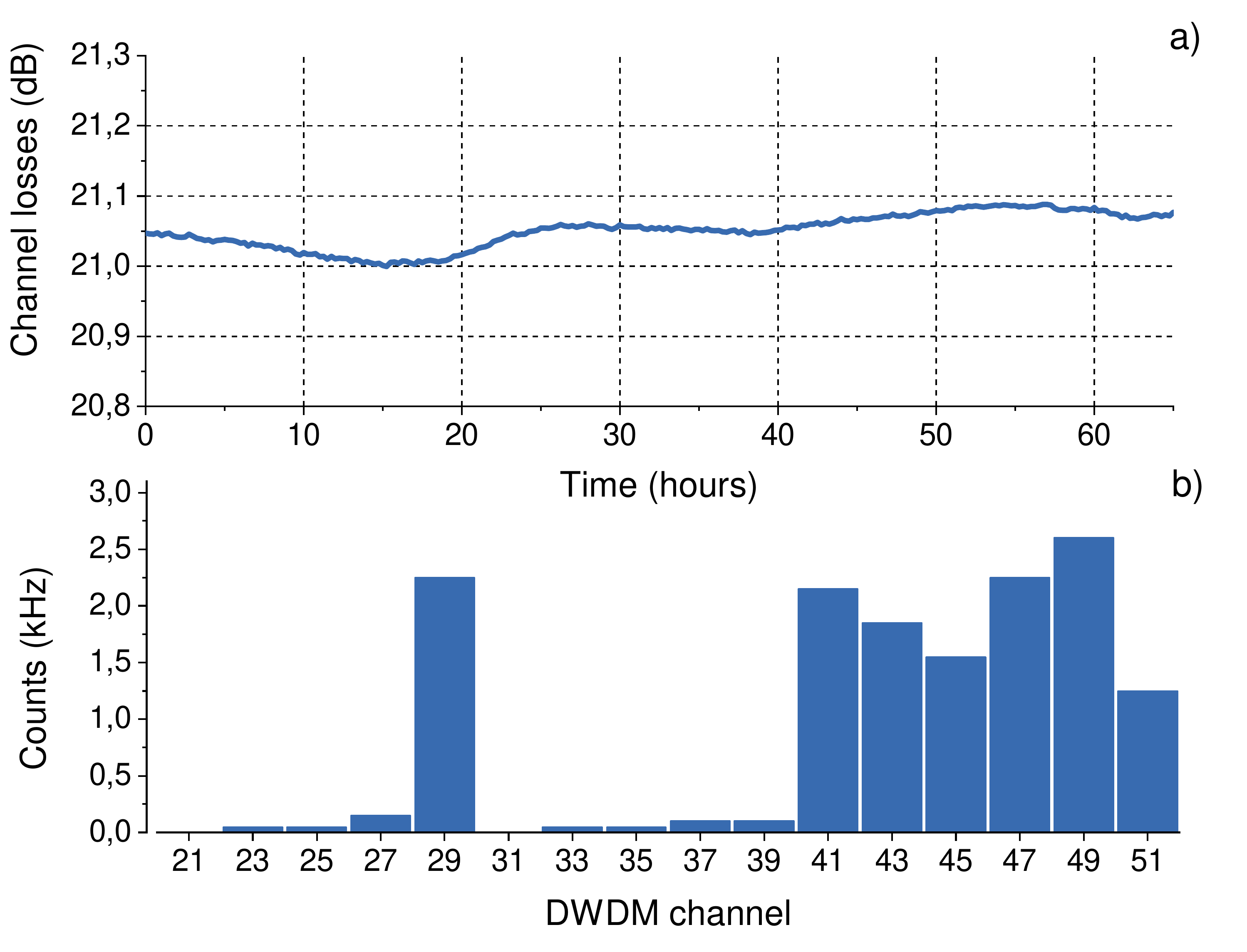}
    \caption{{\bf Characterization of the transmission channel.} a) channel losses during a three-days acquisition; b) noise counts evaluated for different wavelengths with a single photon detector after applying a 200 GHz DWDM filter.}
    \label{fig:stability_cl}
  \end{figure}

\vspace{-0.25cm}
As illustrated in Figure \ref{fig:setup}, the experimental setup consists of a transmitter (Alice) and a receiver (Bob) connected by a metropolitan dark-fiber link in a loop-back configuration. A fiber mirror is installed at one end of the fiber (situated at the telecom datacenter in Florence) in order to drive light back to the European Laboratory for Non-linear Spectroscopy (LENS) where Alice and Bob are located. The total length of the loop-back fiber is about 40 km, with an overall transmission loss of 21 dB. Channel stability in terms of attenuation was monitored for several hours, as reported in Figure \ref{fig:stability_cl} a). 
Figure \ref{fig:stability_cl} b) shows the dark fiber performances in terms of noise, mainly due to interfiber cross-talk. The count rate reported is acquired with a single-photon detector, after applying a 200 GHz dense wavelength division multiplexing (DWDM) filter (detector dark counts are here subtracted, \textit{i.e.} 2700 Hz).\\ 
We performed the three-state BB84 protocol with time-bin encoding in a finite-key scenario, which has the advantage of being a simple and efficient solution for practical QKD \cite{Boaron2018_Simple2.5GHzQKD}. Quantum states belonging to $\mathcal{Z}$ basis are employed for key generation, while $\mathcal{X}$ basis is implemented for security checking (Figure \ref{fig:setup}). The security of this protocol against general attacks has been proven to be maintained, with finite-key analysis, when only two detectors are employed at the receiver, thus simplifying considerably the experimental resources \cite{Rusca2018_SecProofSimpleBB84}. Whenever weak coherent pulses (WCPs) are prepared instead of single photons, a very efficient one-decoy state scheme can be implemented in order to detect photon number splitting attacks \cite{Rusca2018_FiniteKeyAnalysis,lo2005}. This simplified BB84 protocol was demonstrated recently in the record-breaking QKD transmission over 421 km optical fiber \cite{Boaron2018_421km}.\\
To prepare a train of modulated WCPs, Alice carves with an intensity modulator (IMs) the time-encoded pulses from a tunable continuous wave (CW) laser, emitting ITU-T channel 21 ($1560.61$ nm). The repetition rate of quantum states is $\nu = 595$ MHz. A second IM is used to implement the one-decoy state technique with $\mu_1=0.41$ and  $\mu_2=0.15$ photons per pulse (only one intensity is reported in Figure~\ref{fig:setup} for simplicity). 
Then, light is sent through a phase modulator (PM) for phase randomization of each state. An alternative solution would be employing a pulsed laser working in gain switching mode: in this way the phase of the weak coherent states is intrinsically random and the setup can be further simplified. Finally, a variable optical attenuator is used to reach the single photon regime. 
In order to test the QKD system in a more practical scenario, the synchronization signal between users is carried out by classical light propagating through the same fiber together with quantum pulses, by using a 2x1 beam splitter as shown in Fig. 2. To prepare the optical synchronization signal, a second laser working at $1536.61$ nm (DWDM 200 GHz  channel 51) modulated by an extra IM with a custom pattern format at $0.145$ Mbit/s is used. Classical light at -29 dBm launch power is then combined together with quantum pulses and sent through the dark fiber link. 
At Alice's side, four electrical outputs generated by a field programmable gate array (FPGA) are used to drive the IMs for quantum and synchronization signals. Electrical pulse width is approximately $100$ ps, whereas the obtained optical pulse width is around $150$ ps. The PM is driven by a digital-to-analog converter (DAC) which uses 8 bit to obtain $2^{8}-1$ different phase values. Furthermore, a pseudo random binary sequence of $l=2^{12}-1$ bit is used as a key generator, although a quantum random number generator should be used in a real implementation\cite{Avesani2018,Stevanov2000,Jennewein2000}. (This device can be included in future realization directly on Alice's FPGA board.) 
%\section{Results}
%\vspace{-0.3cm}
At Bob's side, a DWDM filter is used to separate classical and quantum light (channel 21 and 51 respectively of the 200 GHz DWDM filter). Classical pulses serve as reference signals for a time tagging unit, which collects the electrical outputs from the two InGaAs single photon detectors (SPDs) that are employed for quantum state measurements \cite{MPD2012}. Bob’s choice of measurement basis is made passively by a 10 dB beam splitter. In the $\mathcal{Z}$ basis, one SPD is used for collecting photons and detecting their arrival time, while pulses measured in the $\mathcal{X}$ basis are sent to an unbalanced fiber-based Mach-Zehnder (MZ) interferometer. A second SPD monitors one of the two outputs of the MZ, in order to check for potential eavesdropping disturbances. 
Additionally, to phase-stabilize Bob's interferometer, a feedback system involving a counter-propagating CW laser (emitting ITU-T channel 35 of a 200 GHz DWDM filter, $1549.32$ nm) is employed. The optical power monitored at the other MZ input allows a piezoelectric system to lock the phase, thus self-stabilizing $\mathcal{X}$ basis measurements.

\section*{Results and discussion}
\vspace{-0.3cm}
\begin{figure}[h!]
      \includegraphics[width=0.465\textwidth, left]{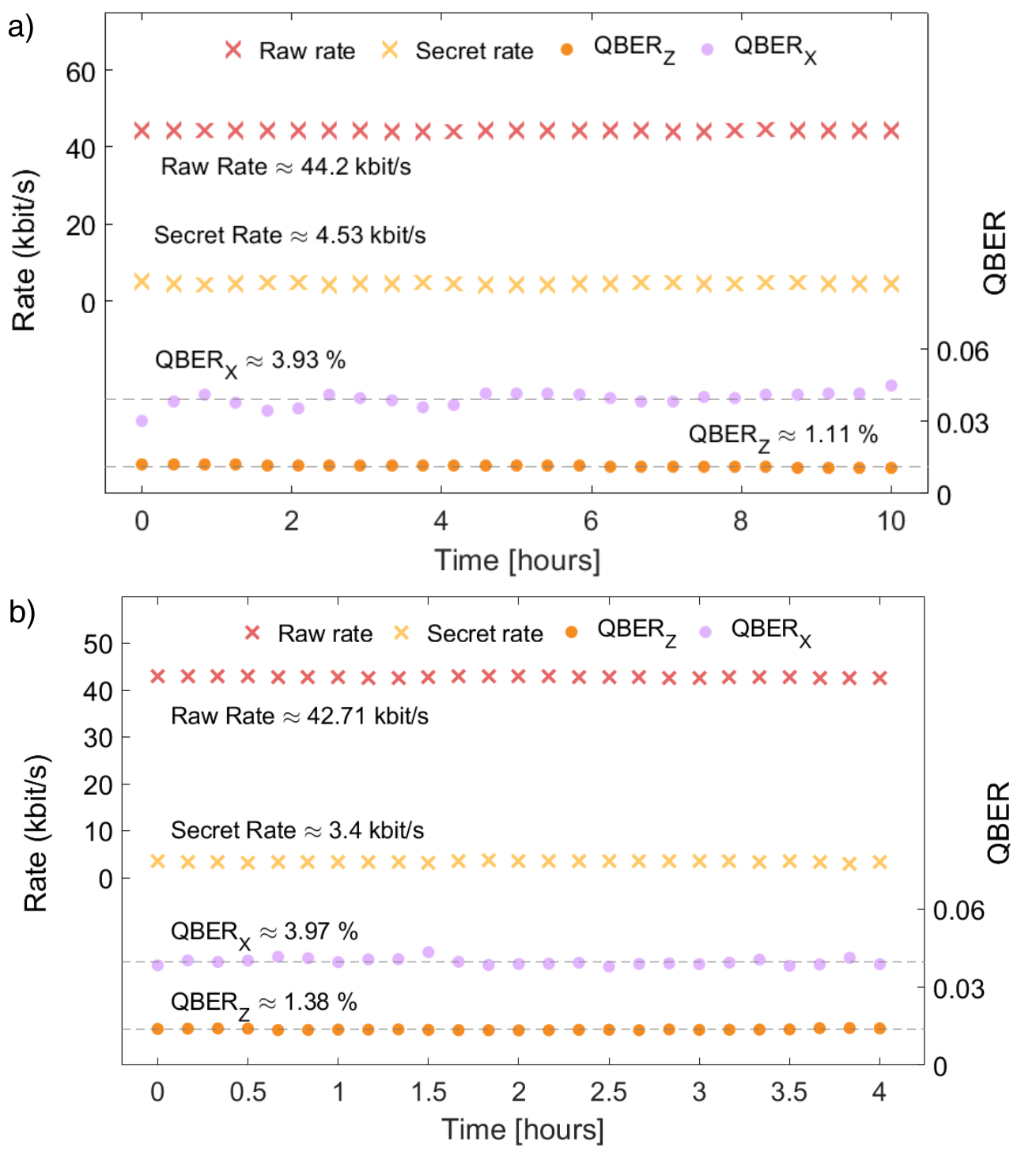}
    \caption{{\bf Stability of the system over time.} Error rates, raw and secret key rate acquired for multiple hours in two different channel conditions: a) only quantum signal in the channel, b) both quantum and classical are copropagating through the fiber.}
    \label{fig:stability_quantum}
  \end{figure}
\begin{figure}[t]
\centering
\includegraphics[width=0.45\textwidth]{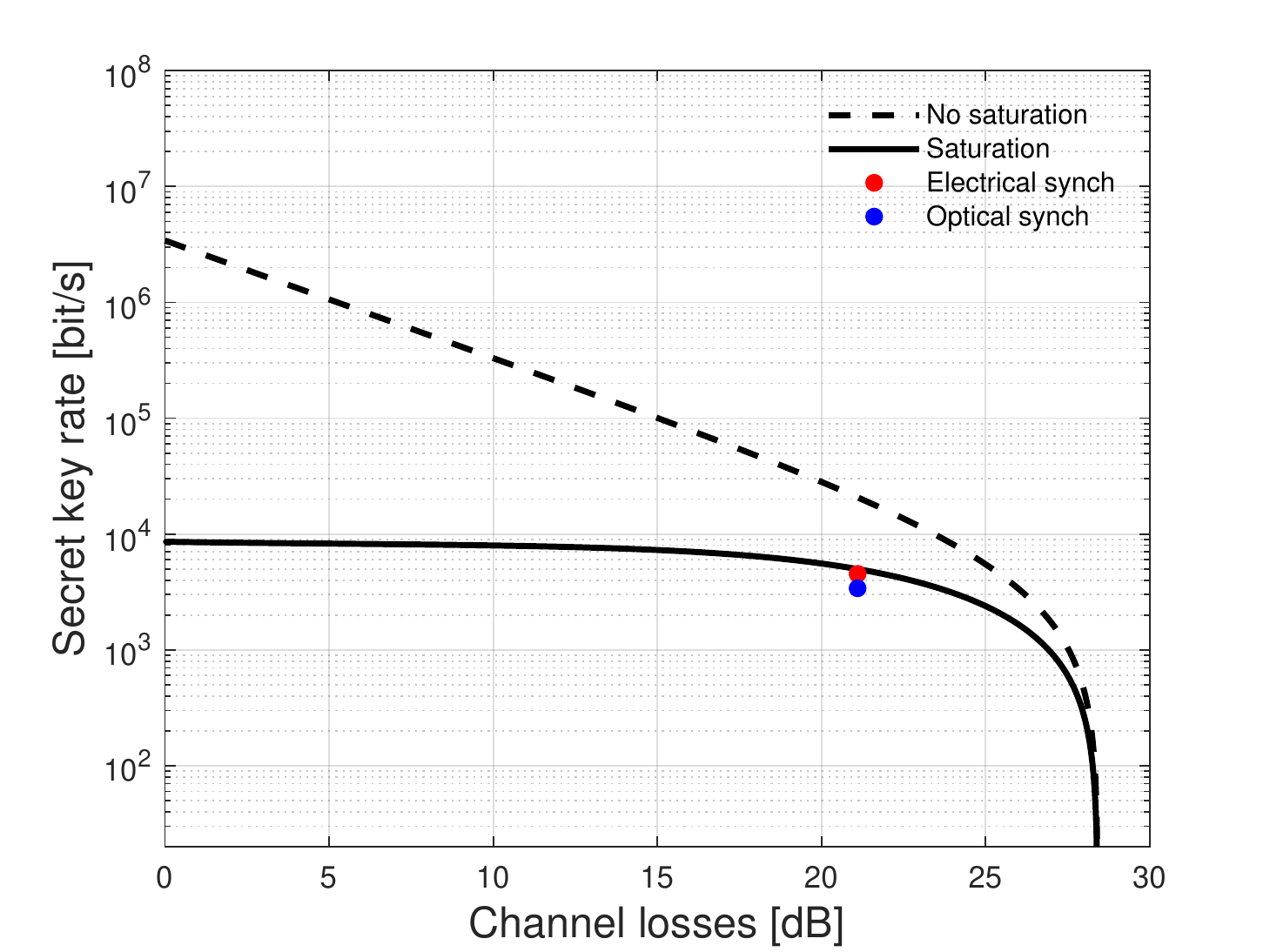}
\caption{{\bf Secret key rate generation.} Secret key rate as a function of the channel losses. The averaged rates achieved in our continuous key exchange for the two cases. The red dot represents the case with only quantum light, whereas the blue dot represents the case with classical and quantum light. The data are averaged over 10 minutes of measurement. Parameters used: probability of $\mathcal{Z}$ basis $p_z=0.9$, probability of $\mathcal{X}$ basis $p_x=0.1$, block size of $n=10^9$, probability of $\mu_1$ and $\mu_2$ of $0.7$ and $0.3$, a secrecy parameter $\epsilon_{sec}$ of $10^{-9}$ with a correctness parameter $\epsilon_{cor}$ of $10^{-9}$, dead time of $20 \ \mu$s, detector efficiency of $20 \%$ and the efficiency of the receiver of $0.5$, and of $0.28$ for $\mathcal{Z}$ basis and $\mathcal{X}$ basis respectively.}
\label{fig:simulation}
\end{figure}

In Figure \ref{fig:stability_quantum} we present the in-field performances of our QKD scheme in terms of quantum bit error rate (QBER) achieved in both measurement bases and final secret key rate (SKR) evaluated with a finite-key analysis \cite{Rusca2018_SecProofSimpleBB84}. We performed a 4 hours continuous key exchange with simultaneous transmission of optical clock signals and a 10 hours continuous key exchange with electrical synchronization provided by Alice's FPGA. These results are achieved without manual stabilization of the experimental setup, thanks to our servo-locking fiber-based interferometer. As expected for our loop-back setup, the average SKR obtained with optical synchronization (3.40 kbit/s) is slightly lower than the average SKR obtained with electrical synchronization (4.53 kbit/s) provided by simply connecting Alice and Bob through the optical bench. This is mainly due to the background noise generated by classical pulses propagating through the same fiber channel, that results in higher QBER in both measurement bases. In addition, electrical synchronization is more stable and accurate than the optical one. A different and more efficient filter can be employed in order to maximize the secret key parameter.\\ 
%\todo[inline]{Am I mistaken or the optical sync should outperform the electrical one once Alice and Bob are widely separated or if an auxiliary fiber is employed?}
As it is shown in Figure \ref{fig:simulation}, our SKR values are compatible with the expected rate achievable by this protocol when single photon detectors operate in saturation regime \cite{Rusca2018_SecProofSimpleBB84}. Furthermore, the simulations show the performances of our system in case of shorter (longer) fiber link with the same intrinsic optical attenuation. Our system can generate a positive key rate up to 28 dB channel losses, which corresponds to 175 km of ultra-low loss single mode fiber.\\
\begin{figure}[h!]
\centering
\includegraphics[width=0.50\textwidth]{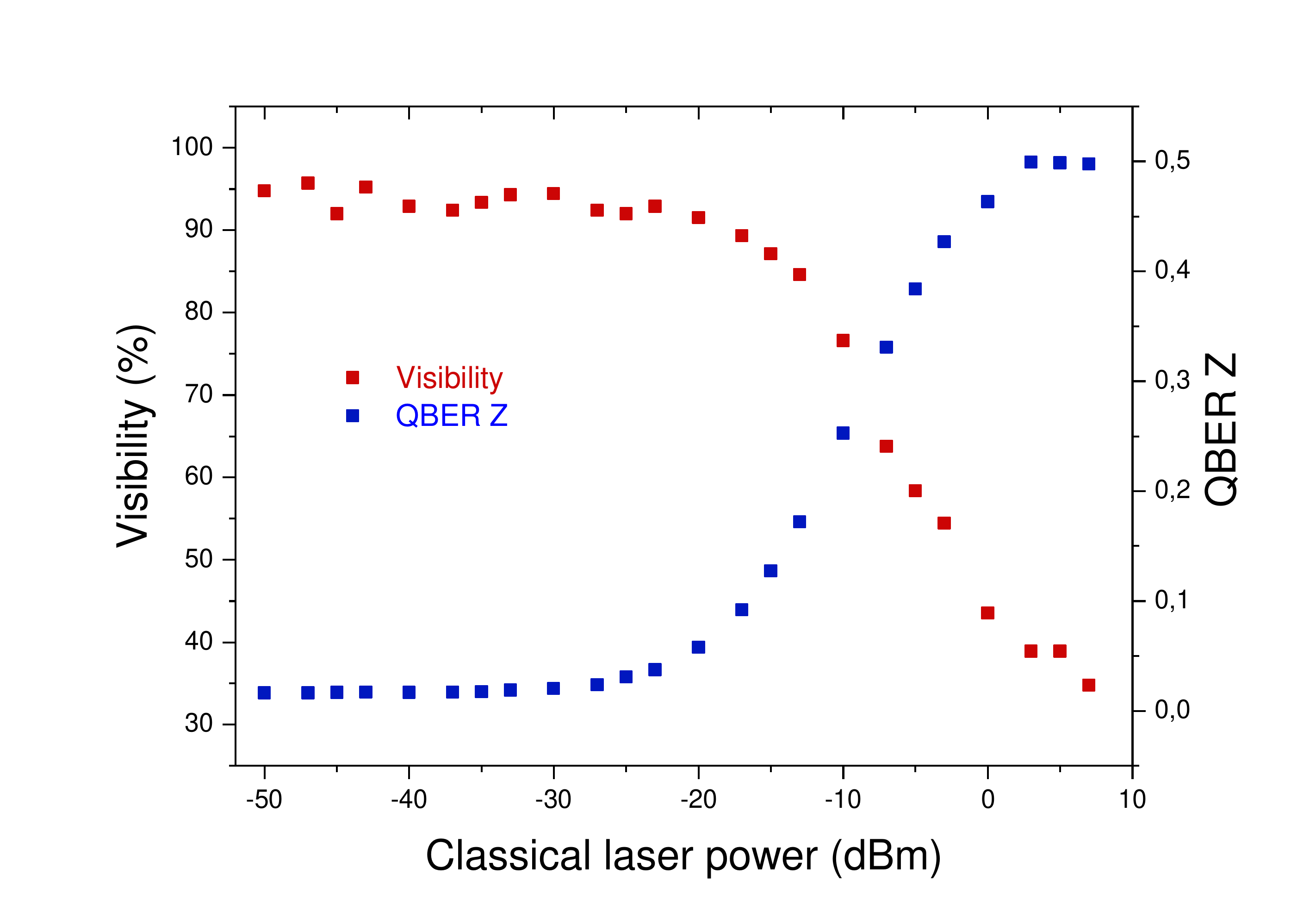}
\caption{{\bf System robustness against classical power.} $\mathcal{Z}$ basis quantum bit error rate and $\mathcal{X}$ basis visibility as functions of classical light power copropagating in the same fiber by means of a DWDM scheme. Secret key rate generation is feasible up to -27 dBm of classical power.}
\label{fig:dataVSclassic}
\end{figure}
%\todo[inline]{Are all the symbols in the caption of Fig.5 so standard that we don't need to explain them (for example $\epsilon_{sec}$ and $\epsilon_{cor}$)? They are quite common, but I will define them in the caption.}
Figure \ref{fig:dataVSclassic} shows the performances of our system (evaluated with electrical synchronization only) with increasing launch power of a CW laser light (emitted at ITU-T Channel 51) sent into the fiber channel, with 21dB loss, together with quantum signals (ITU-T Channel 21). As expected, the increasing classical power degrades both the $\mathcal{Z}$ basis QBER and interferometer visibility (related to $\mathcal{X}$ basis QBER), in a way that at -27 dBm the overall error exceeds the threshold value that prevents a secret key to be shared. At this power level, more advanced multiplexing schemes are required for secret key exchange with this protocol and detectors.

\section*{Conclusions}
In conclusion, we demonstrated a simple and low-cost quantum key distribution system over a field trial link. Quantum and weak classical light have been co-propagated along an installed fiber, proving the stability of the entire system for more than 4 hours. This work acts as a milestone for the future Italian quantum network, proving how the already installed fiber can be pursued for quantum communication over the whole country.  

\section*{Funding}
\vspace{-0.25cm}
 This work is supported by the Center of Excellence, SPOC-Silicon Photonics for Optical Communications (ref DNRF123), by the People Programme (Marie Curie Actions) of the European Union's Seventh Framework Programme (FP7/2007-2013) under REA grant agreement n$^\circ$ $609405$ (COFUNDPostdocDTU). D. Bacco acknowledges financial support from the COST action MP 1403.
This research was sponsored by the NATO Science for Peace and Security program  under grant G5485.

\bibliography{mybib}{}

\begin{thebibliography}{10}

\bibitem{BB84}
C.~H. Bennett and G.~Brassard, ``Quantum cryptography: public-key distribution
  and coin tossing,'' in {\em Proceedings of IEEE International Conference on
  Computers Systems and Signal Processing}, pp.~175--179, 1984.

\bibitem{Scarani2009}
V.~Scarani, H.~Bechmann-Pasquinucci, N.~J. Cerf, M.~Du{\v{s}}ek,
  N.~L{\"u}tkenhaus, and M.~Peev, ``The security of practical quantum key
  distribution,'' {\em Reviews of modern physics}, vol.~81, no.~3, p.~1301,
  2009.

\bibitem{Diamanti2016}
E.~Diamanti, H.-K. Lo, B.~Qi, and Z.~Yuan, ``Practical challenges in quantum
  key distribution,'' {\em npj Quantum Information}, vol.~2, p.~16025, 2016.

\bibitem{Boaron2018_421km}
A.~Boaron, G.~Boso, D.~Rusca, C.~Vulliez, C.~Autebert, M.~Caloz, M.~Perrenoud,
  G.~Gras, F.~Bussi{\`e}res, M.-J. Li, {\em et~al.}, ``Secure quantum key
  distribution over 421 km of optical fiber,'' {\em Physical review letters},
  vol.~121, no.~19, p.~190502, 2018.

\bibitem{Hwang2009}
W.-Y. Hwang, ``Quantum key distribution with high loss: toward global secure
  communication,'' {\em Physical Review Letters}, vol.~91, no.~5, p.~057901,
  2003.

\bibitem{Minder2019}
M.~Minder, M.~Pittaluga, G.~Roberts, M.~Lucamarini, J.~Dynes, Z.~Yuan, and
  A.~Shields, ``Experimental quantum key distribution beyond the repeaterless
  secret key capacity,'' {\em Nature Photonics}, p.~1, 2019.

\bibitem{Yin2016}
H.-L. Yin, T.-Y. Chen, Z.-W. Yu, H.~Liu, L.-X. You, Y.-H. Zhou, S.-J. Chen,
  Y.~Mao, M.-Q. Huang, W.-J. Zhang, {\em et~al.},
  ``Measurement-device-independent quantum key distribution over a 404 km
  optical fiber,'' {\em Physical review letters}, vol.~117, no.~19, p.~190501,
  2016.

\bibitem{Beatrice2018}
B.~Da~Lio, D.~Bacco, D.~Cozzolino, F.~Da~Ros, X.~Guo, Y.~Ding, Y.~Sasaki,
  K.~Aikawa, S.~Miki, H.~Terai, {\em et~al.}, ``Record-high secret key rate for
  joint classical and quantum transmission over a 37-core fiber,'' in {\em 2018
  IEEE Photonics Conference (IPC)}, pp.~1--2, IEEE, 2018.

\bibitem{Dynes2016}
J.~F. Dynes, W.~W. Tam, A.~Plews, B.~Fr{\"o}hlich, A.~W. Sharpe, M.~Lucamarini,
  Z.~Yuan, C.~Radig, A.~Straw, T.~Edwards, {\em et~al.}, ``Ultra-high bandwidth
  quantum secured data transmission,'' {\em Scientific reports}, vol.~6,
  p.~35149, 2016.

\bibitem{Qiu2014}
J.~Qiu, ``Quantum communications leap out of the lab,'' {\em Nature News},
  vol.~508, no.~7497, p.~441, 2014.

\bibitem{Peev2014}
M.~Peev, C.~Pacher, R.~All{\'e}aume, C.~Barreiro, J.~Bouda, W.~Boxleitner,
  T.~Debuisschert, E.~Diamanti, M.~Dianati, J.~Dynes, {\em et~al.}, ``The
  {SECOQC} quantum key distribution network in vienna,'' {\em New Journal of
  Physics}, vol.~11, no.~7, p.~075001, 2009.

\bibitem{Yuan2005}
Z.~Yuan and A.~Shields, ``Continuous operation of a one-way quantum key
  distribution system over installed telecom fibre,'' {\em Optics express},
  vol.~13, no.~2, pp.~660--665, 2005.

\bibitem{Shimizu2014}
K.~Shimizu, T.~Honjo, M.~Fujiwara, T.~Ito, K.~Tamaki, S.~Miki, T.~Yamashita,
  H.~Terai, Z.~Wang, and M.~Sasaki, ``Performance of long-distance quantum key
  distribution over 90-km optical links installed in a field environment of
  {T}okyo metropolitan area,'' {\em Journal of Lightwave Technology}, vol.~32,
  no.~1, pp.~141--151, 2014.

\bibitem{Tang2016}
Y.-L. Tang, H.-L. Yin, Q.~Zhao, H.~Liu, X.-X. Sun, M.-Q. Huang, W.-J. Zhang,
  S.-J. Chen, L.~Zhang, L.-X. You, {\em et~al.},
  ``Measurement-device-independent quantum key distribution over untrustful
  metropolitan network,'' {\em Physical Review X}, vol.~6, no.~1, p.~011024,
  2016.

\bibitem{Bunandar2018}
D.~Bunandar, A.~Lentine, C.~Lee, H.~Cai, C.~M. Long, N.~Boynton, N.~Martinez,
  C.~DeRose, C.~Chen, M.~Grein, {\em et~al.}, ``Metropolitan quantum key
  distribution with silicon photonics,'' {\em Physical Review X}, vol.~8,
  no.~2, p.~021009, 2018.

\bibitem{Collins2016}
R.~J. Collins, R.~Amiri, M.~Fujiwara, T.~Honjo, K.~Shimizu, K.~Tamaki,
  M.~Takeoka, E.~Andersson, G.~S. Buller, and M.~Sasaki, ``Experimental
  transmission of quantum digital signatures over 90 km of installed optical
  fiber using a differential phase shift quantum key distribution system,''
  {\em Optics letters}, vol.~41, no.~21, pp.~4883--4886, 2016.

\bibitem{Zhang2017}
Y.-C. Zhang, Z.~Li, Z.~Chen, C.~Weedbrook, Y.~Zhao, X.~Wang, C.~Xu, X.~Zhang,
  Z.~Wang, M.~Li, {\em et~al.}, ``Continuous-variable {QKD} over 50km
  commercial fiber,'' {\em arXiv preprint arXiv:1709.04618}, 2017.

\bibitem{Tanaka2008}
A.~Tanaka, M.~Fujiwara, S.~W. Nam, Y.~Nambu, S.~Takahashi, W.~Maeda, K.-i.
  Yoshino, S.~Miki, B.~Baek, Z.~Wang, {\em et~al.}, ``Ultra fast quantum key
  distribution over a 97 km installed telecom fiber with wavelength division
  multiplexing clock synchronization,'' {\em Optics express}, vol.~16, no.~15,
  pp.~11354--11360, 2008.

\bibitem{Choi2014}
I.~Choi, Y.~R. Zhou, J.~F. Dynes, Z.~Yuan, A.~Klar, A.~Sharpe, A.~Plews,
  M.~Lucamarini, C.~Radig, J.~Neubert, {\em et~al.}, ``Field trial of a quantum
  secured 10 {G}b/s {DWDM} transmission system over a single installed fiber,''
  {\em Optics express}, vol.~22, no.~19, pp.~23121--23128, 2014.

\bibitem{Wonfor2017}
A.~Wonfor, J.~Dynes, R.~Kumar, H.~Qin, W.~Tam, A.~Plews, A.~Sharpe,
  M.~Lucamarini, Z.~Yuan, R.~Penty, {\em et~al.}, ``High performance field
  trials of {QKD} over a metropolitan network,'' {\em Quantum Cryptography
  (Qcrypt)}, p.~Th467, 2017.

\bibitem{Mao2018}
Y.~Mao, B.-X. Wang, C.~Zhao, G.~Wang, R.~Wang, H.~Wang, F.~Zhou, J.~Nie,
  Q.~Chen, Y.~Zhao, {\em et~al.}, ``Integrating quantum key distribution with
  classical communications in backbone fiber network,'' {\em Optics express},
  vol.~26, no.~5, pp.~6010--6020, 2018.

\bibitem{INRIM2016}
D.~Calonico, ``A fibre backbone in {I}taly for precise time and quantum key
  distribution.'' 4th ETSI/IQC Workshop on Quantum-Safe Cryptography, Toronto
  19-21 Sep 2016.

\bibitem{Boaron2018_Simple2.5GHzQKD}
A.~Boaron, B.~Korzh, R.~Houlmann, G.~Boso, D.~Rusca, S.~Gray, M.-J. Li,
  D.~Nolan, A.~Martin, and H.~Zbinden, ``Simple 2.5 {GH}z time-bin quantum key
  distribution,'' {\em Applied Physics Letters}, vol.~112, no.~17, p.~171108,
  2018.

\bibitem{Rusca2018_SecProofSimpleBB84}
D.~Rusca, A.~Boaron, M.~Curty, A.~Martin, and H.~Zbinden, ``Security proof for
  a simplified {B}ennett-{B}rassard 1984 quantum-key-distribution protocol,''
  {\em Physical Review A}, vol.~98, no.~5, p.~052336, 2018.

\bibitem{Rusca2018_FiniteKeyAnalysis}
D.~Rusca, A.~Boaron, F.~Gr{\"u}nenfelder, A.~Martin, and H.~Zbinden,
  ``Finite-key analysis for the 1-decoy state {QKD} protocol,'' {\em Applied
  Physics Letters}, vol.~112, no.~17, p.~171104, 2018.

\bibitem{lo2005}
H.-K. Lo, X.~Ma, and K.~Chen, ``Decoy state quantum key distribution,'' {\em
  Physical review letters}, vol.~94, no.~23, p.~230504, 2005.

\bibitem{Avesani2018}
M.~Avesani, D.~G. Marangon, G.~Vallone, and P.~Villoresi,
  ``Source-device-independent heterodyne-based quantum random number generator
  at 17 {G}bps,'' {\em Nature communications}, vol.~9, no.~1, p.~5365, 2018.

\bibitem{Stevanov2000}
A.~Stefanov, N.~Gisin, O.~Guinnard, L.~Guinnard, and H.~Zbinden, ``Optical
  quantum random number generator,'' {\em Journal of Modern Optics}, vol.~47,
  no.~4, pp.~595--598, 2000.

\bibitem{Jennewein2000}
T.~Jennewein, U.~Achleitner, G.~Weihs, H.~Weinfurter, and A.~Zeilinger, ``A
  fast and compact quantum random number generator,'' {\em Review of Scientific
  Instruments}, vol.~71, no.~4, pp.~1675--1680, 2000.

\bibitem{MPD2012}
A.~Tosi, A.~Della~Frera, A.~Bahgat~Shehata, and C.~Scarcella, ``Fully
  programmable single-photon detection module for {I}n{G}a{A}s/{I}n{P}
  single-photon avalanche diodes with clean and sub-nanosecond gating
  transitions,'' {\em Review of Scientific Instruments}, vol.~83, no.~1,
  p.~013104, 2012.

\end{thebibliography}
\bibliographystyle{ieeetr}

%\clearpage
%\titleformat*{\section}{\bf\normalsize}

\end{document}